\documentstyle[floats,twocolumn,aps,psfig]{revtex}
\def\abs#1{\vert #1 \vert}

\begin{document}

\draft

\twocolumn[\hsize\textwidth\columnwidth\hsize\csname @twocolumnfalse\endcsname

\title{Magnetic properties of a spin-$\frac{1}{2}$ quadrumer chain}

\author{Andreas Honecker and Wolfram Brenig}

\address{Institut f\"ur Theoretische Physik, TU Braunschweig,
    Mendelssohnstr.\ 3, 38106 Braunschweig, Germany.}

\date{September 19, 2000; revised December 23, 2000}
\maketitle
\begin{abstract}
\begin{center}
\parbox{14cm}{
We study a novel $S=1/2$ cluster chain Hamiltonian which has recently
been proposed in the context of the charge ordered low-temperature
phase of $\alpha'$-NaV$_2$O$_5$. We perform a detailed investigation of
this model within a large range of parameters using perturbation theory
and Lanczos diagonalization. Using model-specific local conservation
laws and parameter-dependent mappings to various effective low-energy
Hamiltonians we uncover a rich phase diagram and several regimes of
gapful spin-excitations. We find that the overall features of recent neutron
scattering data on $\alpha'$-NaV$_2$O$_5$ can be fitted within this
model, however using a set of parameters which seems unlikely.
}

\end{center}
\end{abstract}

\pacs{
\hspace{-2mm}
} 
\vskip1pc]

\section{Introduction}

\begin{figure}[hbt]
\leftline{\psfig{figure=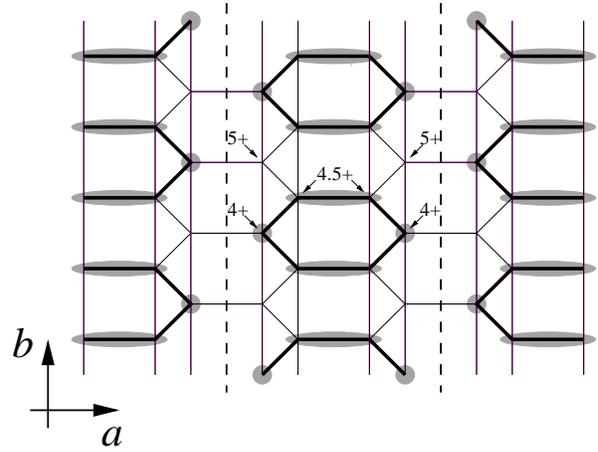,width=0.9\columnwidth}}
\smallskip
\caption{
Possible charge order of Vanadium ions in $\alpha'$-NaV$_2$O$_5$
(schema\-tic). Lines denote the most
important (super)ex\-change paths. The Vanadium atoms are located
at intersections and corners; valence states are indicated by the
numbers in the center of the figure. Grey shaded areas indicate the
localization regime of one unpaired electron.
Decoupling along the dashed lines leads to the quadrumer chain.
\label{figCO}
}
\end{figure}

Low-dimensional magnetic materials exhibit many interesting and
sometimes puzzling properties. In particular, the origin of the spin gap
in the low-temperature phase of $\alpha'$-NaV$_2$O$_5$ is under intense
discussion since it was first reported to open with a phase transition
at $T=34$K \cite{IsUe}. It was confirmed soon \cite{FNYNKKIUS}
that this transition is accompanied with a lattice distortion.
Charge ordering was also realized to play an important r\^ole \cite{OYIU},
leading to various
ordering patterns proposed \cite{CLMcIWG,ThFu,zigzagCO}.
Recently, based on new determinations of the low-temperature
crystal structure \cite{LJSMGK}, two groups \cite{SmLue,BMBP} have
proposed a new type of charge ordering which is sketched in
Fig.\ \ref{figCO}. This structure incorporates Vanadium ions in
three types of valence states: V$^{4+}$, V$^{5+}$ and V$^{4.5+}$.
Entering the low-temp\-era\-ture phase below $T=34$K, every {\em second}
of the quarter-filled two-leg ladders, which make up the system at
high temperatures \cite{SGWPRWG} is in a zig-zag ordered pattern
proposed previously for all of the ladders \cite{zigzagCO}.
The remaining ladders survive the transition.
This new proposal does not completely rule out other scenarios
e.g.\ with only V$^{4+}$ and V$^{5+}$ since the charges were determined only
indirectly through the crystal structure whose determination in itself appears
to be delicate \cite{BMMBP}.
Nevertheless, the charge order proposed by \cite{SmLue,BMBP} is appealing
since it contains clusters of six Vanadium ions with a total of
four unpaired electrons (indicated by bold lines in Fig.\ \ref{figCO})
and therefore a spin gap arises naturally \cite{BMBP}.

\begin{figure}[hbt]
\centerline{\psfig{file=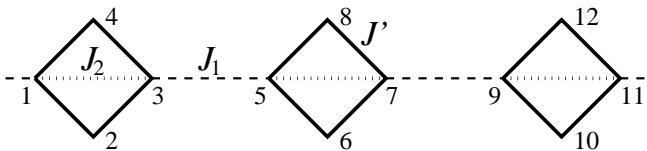,width=\columnwidth}}
\caption{The quadrumer-chain Hamiltonian. $S=1/2$ spins are located
at the numbered intersection points and corners. Coupling constants are
indicated by the style of the lines.
\label{FigMod}
}
\end{figure}

The nature of magnetic excitations above the gap in
the low-temperature phase of $\alpha'$-NaV$_2$O$_5$, {\it i.e.}\
the dispersion of $S=1$ excitations,
has been determined by neutron scattering
\cite{YNKKFIKU,RLBGHCJR,GCRLZBHCJR}: Along the $a$-direction, two
weakly dispersive excitations with an approximate bandwidth of
1meV are observed. In contrast, there is a rather large dispersion
along the $b$-direction \cite{YNKKFIKU}, with a band rising from
around 10meV to at least 40meV \cite{GCRLZBHCJR}.
This suggests that a one-dimensional model extending along the
$b$-axis should result in a reasonable first approximation of the
magnetic properties. In the charge-ordered state of Fig.\ \ref{figCO},
this can be implemented by a decoupling along the dashed lines.
To leading order in $1/U$ of the related Hubbard model, this
leads to the spin-1/2 quadrumer chain of Fig.\ \ref{FigMod} 
\cite{GVAHW}.

The properties of the cluster chain model and its relation to
$\alpha'$-NaV$_2$O$_5$ are subject to controversial discussion
\cite{GVAHW,TrSe}. Moreover, not all regimes of parameters possibly
relevant to $\alpha'$-NaV$_2$O$_5$ have yet been considered in
the literature. In particular, the case $J' < 0$ remains to be investigated
as some LDA+U calculations seem to indicate a negative $J'$ \cite{ThYa}.
Apart from its possible relevance to $\alpha'$-NaV$_2$O$_5$, the quadrumer
model is closely related to several other interesting spin-chain Hamiltonians
\cite{KaIm,RIS,KKKF,KOK}. Eg., CaV$_4$O$_9$ was discussed in the framework
of the model in Fig.\ \ref{FigMod}, however with $J_2 = 0$ \cite{KaIm}. 
The latter model was also generalized \cite{RIS} by adding vertical couplings
in the quadrumers of Fig.\ \ref{FigMod}. {\em Local} reflection symmetry
of these cluster spin models along the chain direction
allows for exact statements regarding the eigenstates and even analytic
determination of the ground state in certain parameter regions \cite{RIS}.
Motivated by the preceding, the main goal of this work is to provide an
additional careful study of the quadrumer spin-chain combining complementary
techniques and enhancing upon the parameter space investigated so far.

The paper is organized as follows: In section \ref{secPrel}, we describe
the quadrumer-chain Hamiltonian more precisely and discuss the limit
of decoupled quadrumers. Then we study the inter-quadrumer coupling
$J_1$ perturbatively in section \ref{secPert}. We proceed with a numerical
analysis of the model in section \ref{secLanc} and try to fit the
neutron scattering data with this method in section \ref{secFit}.
Finally, we summarize and discuss our results in section \ref{secConcl}.
Supplementary results regarding the magnetization process of the
quadrumer-chain model are presented in an appendix.

\section{Preliminaries}

\label{secPrel}

The $S=1/2$ Heisenberg Hamiltonian corresponding to the model in
Fig.\ \ref{FigMod} reads
\begin{equation}
H = \sum_{x=1}^{L/4} \left\{H_{\diamond,x}
 + J_1 \vec{S}_{4x+3} \cdot \vec{S}_{4x+5} \right\}
\label{Hop}
\end{equation}
with
\begin{eqnarray}
H_{\diamond,x} &=& J'
   \left(\vec{S}_{4x+1} + \vec{S}_{4x+3}\right) \cdot
   \left(\vec{S}_{4x+2} + \vec{S}_{4x+4}\right) \nonumber \\
&& + J_2 \vec{S}_{4x+1} \cdot \vec{S}_{4x+3} \, .
\label{Hquadru}
\end{eqnarray}
A more conventional parameterization of the dimerized linear
chain embedded in Fig.\ \ref{FigMod} would be $J_1 = (1-\delta) J$,
$J_2 = (1+\delta) J$. This emphasizes its relation to the
quarter-filled two-leg ladders of the high-temperature phase
\cite{SGWPRWG} whose low-energy properties are equivalent to
spin-1/2 Heisenberg chains. The dimerization is caused by the lattice
distortions accompanying the phase transition \cite{LJSMGK,BMBP}.
Consequently, estimates
for the coupling constant of an effective Heisenberg chain such as
$J \approx 529{\rm K} \approx 45$meV \cite{MMB} can be used
in the high-temperature phase. Such estimates are not expected
to change drastically below the ordering temperature, apart from
the dimerization. In particular
the coupling along the chain should remain antiferromagnetic,
{\it i.e.}\ the physical regime is that with $J_1, J_2 > 0$.
In contrast, the physical regime for $J'$ is less clear. 
Previous investigations of the model (\ref{Hop}) have assumed
$J' > 0$ \cite{GVAHW,TrSe} while, e.g., in \cite{ThYa} a 
{\em ferromagnetic} $J'$ of approximately $-18$meV has been used,
focusing however on a different, {\it i.e.}\ zig-zag, charge order. We
will therefore investigate both signs of $J'$.

As a first step, we consider the case of decoupled quadrumers,
$J_1 = 0$ where the Hamiltonian (\ref{Hop}) decomposes into those
of individual quadrumers (\ref{Hquadru}).
Rewriting the latter as
\begin{equation}
H_{\diamond,x} =
 {J' \over 2} \left( \vec{L}^2_x - \vec{T}_{A,x}^2 - \vec{T}_{B,x}^2\right)
   + {J_2 \over 2} \left(\vec{T}_{A,x}^2 - {3 \over 2}\right)
\end{equation}
with
\begin{eqnarray}
\vec{T}_{A,x} &=& \vec{S}_{4x+1} + \vec{S}_{4x+3} \, , \nonumber \\
\vec{T}_{B,x} &=& \vec{S}_{4x+2} + \vec{S}_{4x+4} \, , \nonumber \\
\vec{L}_x &=& \vec{T}_{A,x} + \vec{T}_{B,x} \, ,
\end{eqnarray}
one infers the spectrum for an isolated quadrumer in
Table \ref{TabQspec}. The case $J' > 0$ was discussed in detail
in \cite{GVAHW}. The ground state is a singlet with a gap to
$S=1$ excitations \cite{BMBP} for sufficiently large $J'$ ($J' > J_2/2$).
For an illustration of the spectrum see \cite{GVAHW}.
The presence of a gap, however, is not immediately clear in two
other regimes, namely for $J' < -J_2$ where the ground state carries
$L_x = 2$ and for $-J_2 < J' < J_2/2$ where it
is degenerate between an $L_x=0$ and $L_x=1$ representation.

\begin{table}[hbt]
\begin{tabular}{ccccc}
$T_{A,x}$ & $T_{B,x}$ & $L_x$ & Eigenvalue & Ground state for ($J_2 > 0$)\\\hline
  0   &   0   &  0  & $-{3 \over 4} J_2$ & $-J_2 < J' < {J_2 \over 2}$ \\
  0   &   1   &  1  & $-{3 \over 4} J_2$ & $-J_2 < J' < {J_2 \over 2}$ \\
  1   &   0   &  1  & ${1 \over 4} J_2$   &  \\
  1   &   1   &  0  & $-2 J' + {1 \over 4} J_2$ & ${J_2 \over 2} < J'$ \\
  1   &   1   &  1  & $- J' + {1 \over 4} J_2$ & \\
  1   &   1   &  2  & $J' + {1 \over 4} J_2$ & $J' < -J_2$ \\
\end{tabular}
\caption{Spectrum of a single quadrumer.
\label{TabQspec}
}
\end{table}

\begin{figure}[hbt]
\centerline{\psfig{file=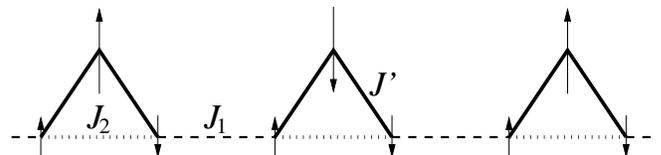,width=\columnwidth}}
\caption{Effective model for the lowest magnetic excitations.
Short arrows denote $S=1/2$ spins and long ones $S=1$. Coupling
constants are indicated by the style of the lines.
\label{FigEffMod}
}
\end{figure}

Even if the coupling $J_1$ is switched on, $T_{B,x}$ remains a good
quantum number {\em locally}, {\it i.e.}\ for all $x$ independently. This is
related to the fact that local reflections of a single quadrumer at
the middle chain are symmetry operations of the model in Fig.\
\ref{FigMod}. Numerical results
(see sections \ref{secLanc} and \ref{secFit} below)
show that the lowest magnetic excitations
are always in the sector with all $T_{B,x} = 1$. These states
are symmetric with respect to all local reflections of quadrumers along
the chain. This finding yields important simplifications in determining
the magnon-dispersion since it suffices to retain only the $S=1$
combination of the two exterior $S=1/2$ spins of a quadrumer. This leads
to an equivalent description in terms of the simplified model sketched
in Fig.\  \ref{FigEffMod}.

\section{Perturbation theory in $J_1$}

\label{secPert}

In this section we focus on a perturbative treatment of inter-quadrumer
coupling. According to the previous section, there are three
regimes with $L_x=0$, 1 and 2 ground states on each quadrumer. While
the gap can be treated by standard non-degenerate perturbation
theory in the case of $L_x=0$, one has to resort to degenerate perturbation
theory in all other cases. Here we show that this leads to effective
Hamiltonians which also have a gap to magnetic excitations. 

\subsection{$J_2/2 < J'$}

For $J_1 = 0$ and $J_2/2 < J'$, the ground state carries
$T_{A,x} = T_{B,x} = 1$, $L_x = 0$. In order to simplify the discussion,
we will also treat $J_2$ perturbatively, {\it i.e.}\ we will
consider the regime $J' \gg \abs{J_1}, \abs{J_2}$. Then the
lowest excited state is obtained by creating an $T_{A,x_0} = T_{B,x_0} = 1$,
$L_{x_0} = 1$ excitation on one quadrumer $x_0$. From this we obtain the
following third-order expansion in $J_1$, $J_2$ of the dispersion
\begin{eqnarray}
{\Delta(k) \over J'} &=& 1 -{\frac {31}{1728}} j_1^{2}
-{\frac {1}{768}} j_1^{3}-{\frac {287}{6912}} j_1^{2}j_2
\nonumber \\
&& +\left ({1 \over 3} j_1+{\frac {7}{72}} j_1^{2}
-{\frac {2825}{165888}} j_1^{3}
+{\frac {13}{432}} j_1^{2}j_2\right )\cos(k) \nonumber \\
&& +\left ({\frac {1}{108}} j_1^{2} + {\frac {31}{1296}} j_1^{3}
+{\frac {5}{162}} j_1^{2}j_2\right )\cos(2 k)
\nonumber \\ &&
+{\frac {13}{1944}} j_1^{3}\cos(3 k)
\label{dispSer}
\end{eqnarray}
with
\begin{equation}
j_i={J_i \over J'} \, .
\end{equation}
Eq.\ (\ref{dispSer}) agrees with the second-order result derived
for $j_2 = 0$ in \cite{KaIm}. One can read off from (\ref{dispSer})
that the minimum of $\Delta(k)$ is at $k=\pi$ for $j_1 > 0$ and
at $k=0$ for $j_1 < 0$. 

Substituting a specific value for $k$ into (\ref{dispSer}),
we can compute two further orders. At $k=0$ and $k=\pi$ we then
find the following fifth-order series for the gap:
\begin{eqnarray}
{\Delta(0) \over J'} &=&
1+{1 \over 3} j_1+{\frac {17}{192}} j_1^{2}+{\frac {403}{20736}}
 j_1^{2}j_2+{\frac {6109}{497664}} j_1^{3} \nonumber \\
&&+{\frac {1657}{248832}} j_1^{2}j_2^{2}+{\frac {229}{36864
}} j_1^{3}j_2-{\frac {5731}{4976640}} j_1^{4} \nonumber \\
&& +{\frac {6763}{2985984}} j_1^{2}j_2^{3}+{\frac {157691}{
71663616}} j_1^{3}j_2^{2}  \nonumber \\
&& -{\frac {29797819}{42998169600
}} j_1^{4}j_2-{\frac {1130744639}{1031956070400}} j_1^{5} \, ,
\label{Delta0Ser} \\
{\Delta(\pi) \over J'} &=&
1-{1 \over 3} j_1-{\frac {61}{576}} j_1^{2}-{\frac {845}{20736}}
 j_1^{2}j_2+{\frac {16403}{497664}} j_1^{3} \nonumber \\
&&-{\frac {6599}{248832}} j_1^{2}j_2^{2}-{\frac {2959}{
995328}} j_1^{3}j_2+{\frac {245833}{9953280}} j_1^{4} \nonumber \\
&& -{\frac {48917}{2985984}} j_1^{2}j_2^{3}+{\frac {
13813}{71663616}} j_1^{3}j_2^{2}  \nonumber \\
&& +{\frac {934164461}{
42998169600}} j_1^{4}j_2-{\frac {7289503141}{1031956070400}} j_1^{5} \, .
\label{DeltaPiSer}
\end{eqnarray}
The series presented above will be compared to numerical results in
section \ref{secLanc} and one will see that they are quite accurate
in the region where they are valid \cite{footnTrSe}.

\subsection{$-J_2 < J' < J_2/2$}

\label{secS1}

For $-J_2 < J' < J_2/2$, one reads off from Table \ref{TabQspec}
that there are two degenerate ground states on each cluster for
$J_1 = 0$. Both
of them carry $T_{A,x} = 0$, one is a singlet with $T_{B,x} = 0$ and the
other one is a triplet with $T_{B,x} = 1$.

\begin{figure}[hbt]
\centerline{\psfig{file=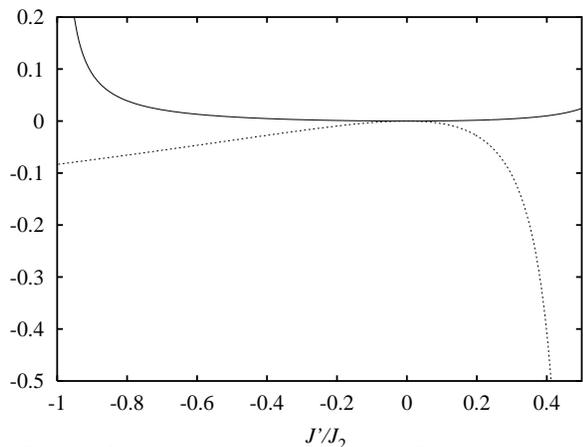,width=\columnwidth,angle=270}}
\caption{
Coupling constants for the effective biquadratic $S=1$ chain:
The lines denote ${\cal J} \times J_2$ (full) and $\beta$ (dotted).
\label{couplFig}
}
\end{figure}

In view of the effective spin-model of Fig.\ \ref{FigEffMod},
only the triplet state with $T_{B,x} = 1$ contributes to the
low-energy magnetic excitations for $J_1 \ne 0$. Based on second-order
degenerate perturbation theory in 
$J_1$ one therefore expects the system to be equivalent to
an effective $S=1$ chain as was already remarked in \cite{GVAHW,TrSe}. 
However, this effective chain is {\em not} a simple $S=1$ Heisenberg-chain:
While first-order interactions vanish, we find already
at second order biquadratic terms which have not been discussed
before. Thus, the effective Hamiltonian is
\begin{equation}
{H^{(S=1)}_{\rm eff.} \over J_1^2} = {e_0 L \over 4}
 + {\cal J} \sum_{x=1}^{L/4} \left \{ \vec{S}_x \cdot \vec{S}_{x+1}
   + \beta \left(\vec{S}_x \cdot \vec{S}_{x+1}\right)^2 \right\} \, ,
\label{HeffS1}
\end{equation}
where $\vec{S}_x$ is an effective $S=1$ operator for the quadrumer $x$.
We have determined the coupling constants in (\ref{HeffS1}) to be
\begin{eqnarray}
e_0 &=& - {\frac {12 \left(J_2 - 3 J'\right) J_2^{3}
+ 17 \left(J' + J_2\right) J_2 {{J'}}^{2} - 8 {{J'}}^{4}}
  {32 \left ({J_2}-2 {J'}\right )
\left (2 {J_2}-{J'}\right )\left (2 {J_2}-3 {J'}\right )
\left ({J_2}+{J'}\right ){J_2}}} \, , \nonumber \\
{\cal J} &=& {\frac {\left (4 {J_2} - {J'}\right ) {{J'}}^{2}}
{32 \left (2 {J_2}-{J'}\right )
 \left (2 {J_2}-3 {J'}\right )\left ({J_2}+{J'}\right ){J_2}}}
\, , \nonumber \\
\beta &=&
- {\frac {\left (3 {J_2}-2 {J'}\right ){{J'}}^{2}}
{2 \left ({J_2}-{J'}\right )\left (4 {J_2} - {J'}\right )
\left ({J_2} - 2 {J'}\right )}} \, .
\label{effS1coupl}
\end{eqnarray}
These effective coupling constants are shown in Fig.\ \ref{couplFig}
for the region where the mapping applies. One finds that ${\cal J}$
is always positive (antiferromagnetic) while $\beta$ is always
negative. At the boundaries of the figure, we have the following
divergences: ${\cal J} \to \infty$ as $J' \to -J_2$ and
$\beta \to - \infty$ as $J' \to J_2/2$. These divergences
are expected since they signal the limit of validity of the mapping to
an effective $S=1$ chain.

The effective Hamiltonian (\ref{HeffS1}) with coupling constants
(\ref{effS1coupl}) is gapful
with the exception of $J' = 0$ and $J' \approx 0.4466 J_2$
(corresponding to $\beta=-1$ -- see for instance Sec.\ 6 of \cite{Affleck}).
The ground state is the Haldane phase for $-J_2 < J'\lesssim 0.4466 J_2$
and it is spontaneously dimerized only in the small window
$0.4466 J_2 \lesssim J' < J_2/2$ (see Sec.\ 6 of \cite{Affleck}).
Interestingly though, the latter window
includes the region presumably most relevant
to $\alpha'$-NaV$_2$O$_5$, which we will find to be $J' \approx J_2/2$
(see section \ref{secFit} below). The biquadratic term thus has the
important consequence that the relevant parameter region of the quadrumer
chain is not adjacent to a Haldane state, but to a dimerized state.

Regarding the size of the gap obtained by this mapping,
one observes that ${\cal J}$ is very small
in most of the region covered by Fig.\ \ref{couplFig}. For example,
for $-0.7264 J_2 < J' < J_2/2$, we have $J_2 \times {\cal J} < 1/40$.
In combination with the additional factor $J_1^2$ this leads to an
extremely small gap of the original Hamiltonian at least in the region
$J_1 \ll J_2$ where the perturbative approach is valid.

\subsection{$J' < -J_2$}

Finally, for $J' < -J_2$ and $J_1 = 0$, the ground state of each quadrumer
is an $L_x=2$ representation. Using first-order degenerate perturbation
theory in $J_1$, the quadrumer chain can then be mapped to an effective
$S=2$ Heisenberg chain
\begin{equation}
H^{(S=2)}_{\rm eff.} = {J_1 \over 16} \sum_{x=1}^{L/4}
     \vec{S}_x \cdot \vec{S}_{x+1} \, ,
\label{HeffS2}
\end{equation}
where now $\vec{S}_x$ is an effective $S=2$ operator for the quadrumer $x$.
The gap of the effective $S=2$ chain was
estimated by DMRG to be $\Delta_{\rm eff.} = (0.0876 \pm 0.0013) J_{\rm eff.}$
\cite{WQY}. Substitution of $J_{\rm eff.} = J_1/16$ into this
DMRG result then leads to a small, but non-zero gap of the quadrumer model
\begin{equation}
\Delta \approx 0.0055 J_1 + {\cal O}(J_1^2)
\label{GapExpS2}
\end{equation}
for $J' < -J_2$ and weakly antiferromagnetic $J_1 > 0$.

\section{Lanczos diagonalization}

\label{secLanc}

We now proceed to study the spin gap over a wider range of parameters
using Lanczos diagonalization. We have used periodic boundary
conditions and mostly worked with the original model (\ref{Hop}),
only for the case of $L=32$ spin-1/2 spins have we resorted
to the effective Hamiltonian of Fig.\ \ref{FigEffMod} to
reduce the dimensionality of the Hilbert space.

\begin{figure}[hbt]
\centerline{\psfig{file=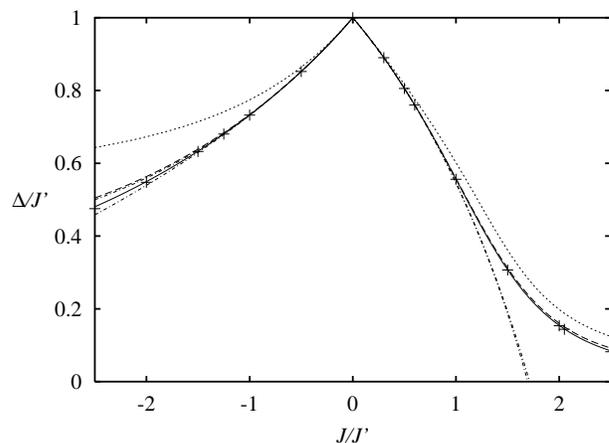,width=\columnwidth,angle=270}}
\caption{
Spin gap as a function of $J = J_1 = J_2$ for $J' > 0$.
Lines are for $L=8$ (dotted), $L=16$ (dashed) and $L=24$ (full).
The symbols `$+$' denote an extrapolation to the thermodynamic limit
and the dashed-dotted lines
[2,3] and [3,2] Pad\'e approximants to the series (\ref{Delta0Ser}) for
$J < 0$ and (\ref{DeltaPiSer}) for $J > 0$.
\label{gapAF}
}
\end{figure}

We begin with parameters that permit a comparison with perturbative
results of the previous section. Fig.\ \ref{gapAF} shows the spin gap,
{\it i.e.}\ the minimum of the one-magnon dispersion, as a function of
$J$ for $J' > 0$ and a homogeneous chain with $J=J_1=J_2$, including the
region $\abs{J} \ll J'$ where perturbation theory should be accurate. In this
figure, we show numerical data for $L=8$, $16$ and $24$.  Actually, we have
determined the lowest excitations in the $S^z = 1$ sector, but we are confident
that this is in fact always an excitation with total spin $S=1$. The
finite-size data was extrapolated with a Shanks transformation \cite{Shanks},
yielding the pluses in Fig.\ \ref{gapAF}. These extrapolations are
indistinguishable from the $L=24$ data, showing that finite-size effects
are extremely small at least for the parameters covered by the figure.

Recall that
eq.\ (\ref{dispSer}) shows that the minimum of the one-magnon dispersion
is located at $k=0$ for $J<0$ and at $k=\pi$ for $J > 0$. This is confirmed
by our numerical data. Therefore, one should compare the numerical data 
to the series (\ref{Delta0Ser}) and (\ref{DeltaPiSer}) for $J<0$ and $J>0$,
respectively. Pad\'e approximants to these series are shown in
Fig.\ \ref{gapAF} and one observes good agreement for $\abs{J} \lesssim J'$.
While the approximants stay close to the numerical data for negative
$J$ as large as $J = -2.5 J'$, systematic deviations are observed at
large positive $J$. This can be attributed to the fact that other
$S=1$ excitations start to mix in at $J' \approx J$ (compare Table
\ref{TabQspec} and \cite{TrSe}).

As a first summary, we find good agreement between our numerical data
and perturbation theory in the regime where the latter should be accurate.
For $J \gtrsim J'$, however, the numerical approach is far
superior since the nature of the lowest $S=1$ excitation becomes
more complicated. Yet, finite-size effects still remain small.

\begin{figure}[hbt]
\centerline{\psfig{file=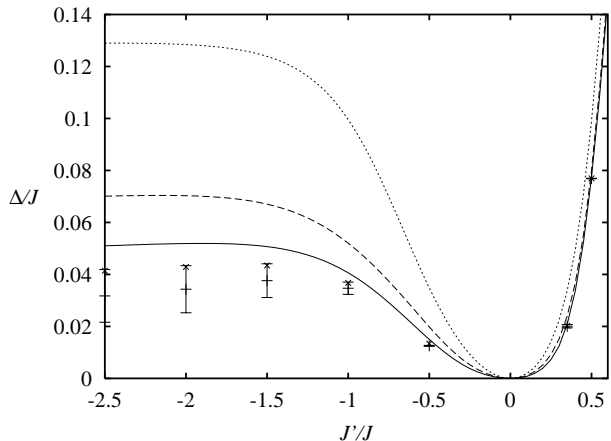,width=\columnwidth,angle=270}}
\caption{
Same as Fig.\ \ref{gapAF}, but now as a function of $J'$ for $J > 0$.
The symbols `$\times$' denote $L=32$ data and a `$+$' with error bars
an extrapolation to the thermodynamic limit.
\label{gapF}
}
\end{figure}

We now concentrate on the region $J > 0$, {\it i.e.}\ the one
which is appropriate for $\alpha'$-NaV$_2$O$_5$, and study the
dependence on $J'$, focusing on small and negative values. 
Fig.\ \ref{gapF} again shows the spin gap, now
as a function of $J'$ for $J = J_1 = J_2 > 0$.
The extrapolations have again been performed with a Shanks transformation
\cite{Shanks}. The actual estimate is based on the $L=16$, $24$ and
$32$ data with error bars determined from the difference with respect to
an extrapolation based on $L=8$, $16$ and $24$. For $J' > 0$,
the trend observed in Fig.\ \ref{gapAF}, {\it i.e.}\ that
$L=8$, $16$ and $24$ is completely sufficient for the extrapolation, is
confirmed.
For $J' < 0$ and small, the errors first remain small with an increasing
uncertainty as $J'$ becomes large and negative. Larger error bars
indicate larger finite-size effects which in turn originate
from an increasing correlation length.

Fig.\ \ref{gapAF} shows that the gap vanishes at $J' = 0$
at any finite system size, the reason being that the exterior
spins are free in this case and give rise to excitations with
exactly zero energy. For any non-zero $J'$, in contrast, the gap seems
to remain non-zero even in the thermodynamic limit.

A maximum of $\Delta/J$ emerges in the ferromagnetic region
$-2 J \lesssim J' \lesssim -J$. In fact, while $\Delta/J$ is already
of the correct order of magnitude, it still has
to decrease substantially for $J' \to -\infty$ in order to reach
the value given by (\ref{GapExpS2}).
In the limit $J' \to -\infty$, we also have to recover the correlation
length $\xi$ of the $S=2$ chain which is known to be $\xi \approx 50$
clusters ({\it i.e.}\ $L \approx 200$ -- see for example \cite{KGWB}),
thus leading to the increasing finite-size effects in Fig.\ \ref{gapF}
as $J'$ goes to larger negative values.

In the region of positive $J,J' > 0$, our results can be compared to DMRG
results (see Fig.\ 2 of \cite{GVAHW}). While we find agreement for large $J'$,
deviations of the order of $J/50$ can be observed in the region of small $J'$.
For instance, we extrapolate a gap $\Delta \approx 0.077 J$ for $J' =0.5J$
while the corresponding value of \cite{GVAHW} is $\Delta \approx 0.05 J$.
Furthermore, the DMRG result of \cite{GVAHW}
for the gap is essentially zero already for $J' = 0.4J$
while our extrapolation at $J' = 0.35 J$ yields $\Delta = 0.0201(6)J$
which is clearly non-zero. This discrepancy can be observed already at
a fixed system size (e.g.\ $L=32$) \cite{Gpriv}, {\it i.e.}\ it is not
due to the extrapolation. There are two main
distinctions between our Lanczos and the DMRG results of \cite{GVAHW}:
First, DMRG is subject to truncation errors which are absent in our approach.
Second, we use periodic boundary conditions in contrast to open ones
in \cite{GVAHW}. Periodic boundary conditions typically have smaller
finite-size effects than open ones and open boundary conditions can even
lead to further boundary excitations. We therefore believe that extrapolation
of our Lanczos data for system sizes up to $L=32$ yields the most accurate
results obtained so far for the thermodynamic limit of the model (\ref{Hop}).

\section{Fitting the neutron scattering data}

\label{secFit}

We have seen so far that there are several regions where different mechanisms
open a spin gap in the quadrumer-chain model with a large variation of its
size. While this looks promising for the applicability of the model
to $\alpha'$-NaV$_2$O$_5$, the real test is whether the neutron-scattering
results for the dispersion along the $b$-axis \cite{YNKKFIKU,GCRLZBHCJR} can
be fitted with reasonable parameters.

Here we follow \cite{GVAHW} and use the following criteria: (i) The
low-temperature experimental data \cite{YNKKFIKU,GCRLZBHCJR} has a miminum
in the center of the high-temperature Brillouin zone. Taking into
account doubling of the unit cell (as is appropriate for the quadrumer-chain
model), the minimum of the dispersion should therefore be located at
$k = 2 \pi \equiv 0$ and should have a value $\Delta(0) \approx 10$meV.
(ii) Since the magnon dispersion was recentely traced up to energies of
$40$meV \cite{GCRLZBHCJR}, one should have $\Delta(\pi) \ge 40$meV.

\begin{figure}[hbt]
\centerline{\psfig{file=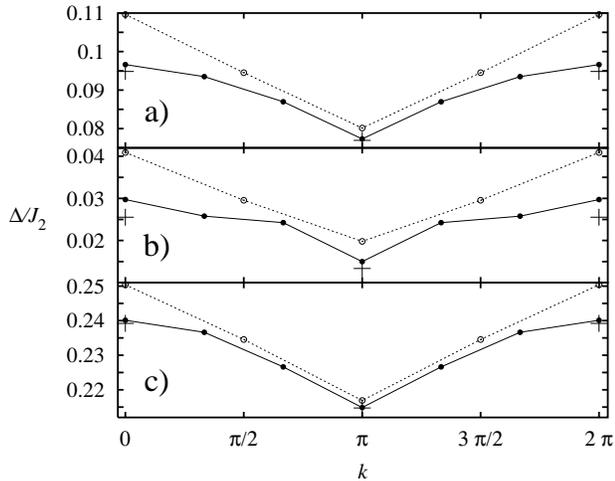,width=\columnwidth,angle=270}}
\caption{
Dispersion of lowest $S = 1$ excitation for 
a) $J_1 = J_2 = 2 J'$,
b) $J_1 = J_2 = - 2 J'$ and
c) $J_1 = 0.9 J_2$, $J' = 0.65 J_2$.
Symbols are for $L=16$ (open circles), $L=24$ (filled circles) and $L=32$
(pluses). Lines are guides to the eye.
\label{DispFig}
}
\end{figure}

In Figs.\ \ref{DispFig}a),b) we first show the dispersion of the lowest
$S=1$ excitation corresponding to the two points $J' = \pm J/2$
in Fig.\ \ref{gapF}. One observes that the minimum is located at $k=\pi$
and remains there after extrapolation to the thermodynamic limit.
Fig.\ \ref{DispFig}c) shows the dispersion for $J' = 0.65J_2$ and
a small dimerization
$\delta \approx 0.053$ which is close to a parameter set proposed
as a possible fit for $\alpha'$-NaV$_2$O$_5$ in \cite{TrSe}. One
observes that our data is already well converged with system size and
one can thus easily extrapolate it to $L = \infty$. We find
$\Delta(0) \approx 0.239 J_2$, which agrees with \cite{TrSe},
but the minimum is still at $k=\pi$: $\Delta(\pi) \approx 0.215 J_2$,
leading to $\Delta(\pi)/\Delta(0) \approx 0.9$.
We therefore agree with \cite{GVAHW} that this parameter region is
{\it not} appropriate for $\alpha'$-NaV$_2$O$_5$. 

\begin{figure}[hbt]
\centerline{\psfig{file=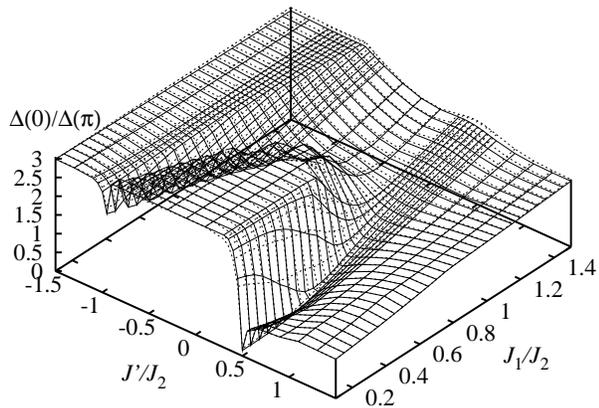,width=\columnwidth,angle=270}}
\caption{
$\Delta(0)/\Delta(\pi)$ as a function of coupling to external spins
$J'/J_2$ and coupling between quadrumers $J_1/J_2$ for $L=16$
(dotted lines) and $L=24$ (full lines).
\label{scanFig}
}
\end{figure}

In order to test whether a good fit can be obtained at all, we
have performed a systematic numerical scan of $\Delta(0)/\Delta(\pi)$
at $L=16$ and $L=24$. The range of interest including
ferromagnetic couplings $J'<0$ we find, is depicted in 
Fig.\ \ref{scanFig}. Apart from this range we have also studied a wide
region of $J'/J_2$ and $J_1/J_2$. We found little variation outside the
parameter window shown in Fig.\ \ref{scanFig}. Note that $\Delta(k)$ was
determined for $S^z = 1$ excitations
which do not necessarily carry $S=1$, but may have higher spin.
Furthermore, there should be a two-particle state at $k=0$
with an energy not larger than $2 \Delta(\pi)$ in the thermodynamic limit.
The ratio $\Delta(0)/\Delta(\pi)$ should therefore not exceed
a value of $2$ in the thermodynamic limit unless the gap closes. This indicates
large finite size effects in some regions of Fig.\ \ref{scanFig}.
However, we are interested in parameters where $\Delta(0)/\Delta(\pi) < 1$.
In that case no complications due to multi-magnon excitations are
expected.

\begin{figure}[hbt]
\centerline{\psfig{file=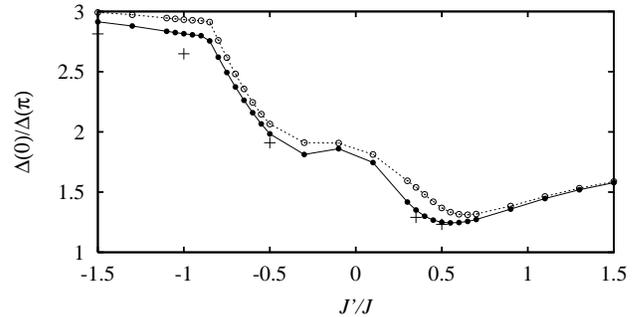,width=\columnwidth,angle=270}}
\caption{
$\Delta(0)/\Delta(\pi)$ for $J_1 = J_2 = J$ as a function of $J'$
for $L=16$ (open circles), $L=24$ (filled circles) and $L=32$
(pluses). Lines are guides to the eye.
\label{scanOneFig}
}
\end{figure}

For clarity, Fig.\ \ref{scanOneFig} shows a section along $J_1 = J_2$ of
Fig.\ \ref{scanFig} including some $L=32$ data points (compare also
Figs.\ \ref{DispFig}a),b)).
One observes that finite-size effects can be important, in particular for
$J' < 0$ where $\Delta(0)/\Delta(\pi)$ exceeds the aforementioned 
limiting value of two. It is evident from this figure that finite-size
effects reduce $\Delta(0)/\Delta(\pi)$. The condition
$\Delta(0) < \Delta(\pi)$ appears nevertheless
impossible to satisfy at $J_1 = J_2$. Moreover, in the entire region
with $J' < 0$, Fig.\ \ref{scanFig} clearly shows that $\Delta(0)$ is always
larger than $\Delta(\pi)$, restricting the possible parameters for a fit to
lie in the region $J' > 0$. This is consistent with refs.\ \cite{GVAHW,TrSe}.

\begin{figure}[hbt]
\centerline{\psfig{file=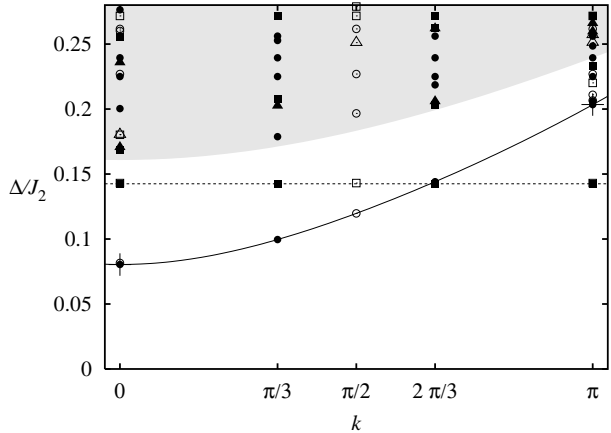,width=\columnwidth,angle=270}}
\caption{
Low-lying excitations at $J_1/J_2 = 7/20$, $J'/J_2 = 11/20$.
Open symbols denote data for $L=16$ and filled symbols for $L=24$.
Boxes stand for $S=0$ excitations, circles for $S=1$ excitations
and triangles for $S=2$ excitations. The pluses show the lowest
$S=1$ excitations for $L=32$ at $k=0$ and $\pi$.
The full line denotes the lowest $S=1$ excitation (magnon) whereas
the dashed line shows the lowest, non-dispersive $S=0$
excitation. The shaded area indicates the expected location of
the two-magnon continuum.
\label{dispVal}
}
\end{figure}

On the other hand, there is a narrow region around $J'= J_2/2$
with $J_1 < J_2$ where $\Delta(0)$ becomes smaller than
$\Delta(\pi)$. An example of an excitation spectrum in this region is
shown in Fig.\ \ref{dispVal}. One does indeed see a single magnon
excitation with a minimum at $k=0$ whose dispersion is interpolated
by the full line. This dispersion would be consistent
with early neutron scattering data \cite{YNKKFIKU}. However, very recent
measurements  \cite{GCRLZBHCJR} clearly show that the
bandwith in Fig.\ \ref{dispVal} is too small. One may obtain
a larger bandwith within the present model by changing the coupling
constants slightly, however, only at the expense
of reducing $\Delta(0)$. In order to have $\Delta(0) = 10$meV and
$\Delta(\pi) \ge 40$meV one then needs a coupling constant
$J = J_2/(1+\delta) \gtrsim 300$meV with a dimerization
$\delta \approx 0.38...0.54$ and $J' \approx (0.5 \ldots 0.55) J_2$
\cite{footn2}. Both, the large values of $J$ and of the dimerization
are certainly too large to be plausible (see section \ref{secPrel}).
Therefore, even though our numerical results differ in detail we agree
with \cite{GVAHW} regarding that the cluster spin model (\ref{Hop})
does not result in a quantitive fit for $\alpha'$-NaV$_2$O$_5$ with
plausible parameters.

Among the remaining excitations displayed in Fig.\ \ref{dispVal}, there is
a localized ($k$-independet) singlet with $\Delta(k) \approx 0.14J_2$
(indicated by the dashed line).
This excitation can be interpreted as a singlet $T_{B,x_0}=0$ in one quadrumer
$x_0$ (while all other $T_{B,x} = 1$) which is prevented from propagating
since $T_{B,x}$ is locally conserved. Note that such excitations are not
present in the effective model of Fig.\ \ref{FigEffMod}, but only in the
original one of (\ref{Hop}).

We would like to conclude this section with a remark on Fig.\ \ref{scanFig}:
There is one dip in $\Delta(0)/\Delta(\pi)$ for $J_1 < J_2$ both for
negative and positive $J'$. These dips become sharp for $J_1 \to 0$
where they signal transitions between different local ground states.
In this limit, the dips are located at $J'/J_2 = -1$ and
$J'/J_2 = 1/2$ where according to Table \ref{TabQspec} the transitions
occur at $J_1 = 0$. In general, our numerical data is consistent with
a vanishing gap at the location of these dips, {\it i.e.}\ a continuous
transition between different phases. However, due to the large correlation
lengths expected in some of the phases, larger system sizes would be
needed in order to determine the phase diagram of the model (\ref{Hop})
accurately.

\section{Conclusions and outlook}

\label{secConcl}

We have studied an $S=1/2$ quadrumer-chain model
using perturbation theory and the Lanczos method. This model has
been proposed to explain the magnetic excitations of $\alpha'$-NaV$_2$O$_5$
in its low-temperature phase \cite{SmLue,BMBP,GVAHW}. We have shown the
quadrumer-chain Hamiltonian to display a rich phase diagram which we believe
the present study has just begun to unveil (compare Fig.\ \ref{scanFig}):
Different phases are smoothly connected to the local $L_x = 0$, $1$ and $2$
ground states at $J_1 = 0$. In addition, the mapping to an effective $S=1$ chain
in section \ref{secS1} shows that the $L_x = 1$ region consists of a
Haldane and a dimerized phase at least at small $J_1$. The latter
phase arises because of a biquadratic interaction which has not been
realized in previous studies. However, further work is needed to determine
all of the phase boundaries of the quadrumer chain accurately.

The model (\ref{Hop}) gives also rise to interesting behavior in
an external magnetic field. The case $J_1 = J_2$ is discussed
in an appendix where we show that in addition to a spin gap one
also finds a plateau in the magnetization curve at half the saturation
magnetization. This is similar as for the model studied in \cite{KOK}, though
it remains to be investigated whether one can also obtain
other plateaux in the present model for $J_1 \ne J_2$ like the one at
a quarter of the saturation magnetization found in \cite{KOK}.

Finally, we have assessed the relevance of the quadrumer-chain model
to the magnetic excitations of $\alpha'$-NaV$_2$O$_5$. We found that
in order to fit the neutron scattering data for the dispersion
along the chain direction \cite{GCRLZBHCJR} one has to resort
to parameters which are not very plausible. A good fit can even
be obtained for $J_i < 0$, (compare (\ref{dispSer}) and Fig.\ \ref{gapAF})
although ferromagnetic exchange along the chain is clearly an unphysical
choice. The applicability of the quadrumer model to $\alpha'$-NaV$_2$O$_5$
may therefore be questioned \cite{GVAHW}, leaving the proper microscopic model
for this material an open issue. In fact, other models proposed in this context
are similarly deficient. Assuming, e.g., pure zig-zag charge order
\cite{zigzagCO}, the lowest $S=1$ excitation would belong to a two-particle
continuum (see, e.g., \cite{ShSu}) rather than the experimentally observed
magnon state \cite{YNKKFIKU,GCRLZBHCJR}. A possible remedy is to attribute the
spin gap to dimerization \cite{GrVa} rather than frustration. However, this
proposal has been disputed \cite{TrSe} based on the spatial symmetries
of the charge ordered state. Clearly, further experimental
input is needed to decide these issues. For the moment, we cannot rule out
that the correct description of $\alpha'$-NaV$_2$O$_5$ will turn out to be a
modification of the cluster model discussed in the present work.

\acknowledgements

We would like to thank T.\ Chatterji and C.\ Gros for relevant
discussions and correspondence. This work has benefited from
the Schwerpunkt 1073 of the Deutsche Forschungsgemeinschaft.

\appendix

\section*{Magnetization process}

\begin{figure}[hbt]
\centerline{\psfig{file=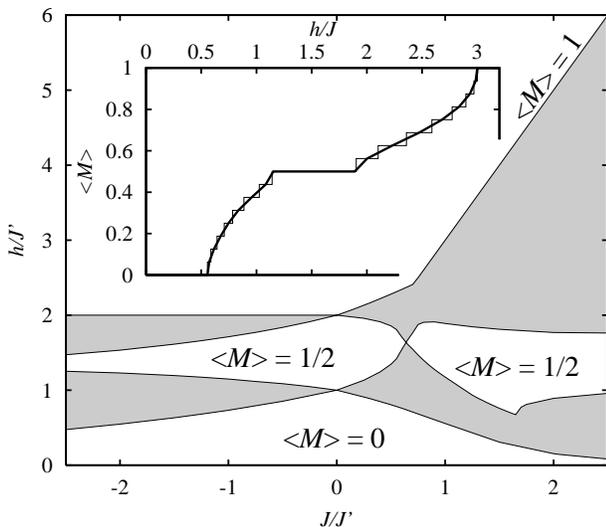,width=\columnwidth,angle=270}}
\caption{
Groundstate phase diagram in a magnetic field $h$ for $J_1 = J_2 = J$.
White areas denote parameter regions with magnetization plateaux
with the indicated values of the magnetization $\langle M \rangle$ whereas the
grey shaded regions denote smooth transitions.
Inset: Magnetization curve at $J' = J$ for $L=32$ (thin
line) and an extrapolation to the thermodynamic limit (bold line).
\label{mpdFig}
}
\end{figure}

In this appendix we discuss the magnetization process of the quadrumer-chain
model at $J_1 = J_2 = J$. Application of an external magnetic field amounts
to adding a term $-h \sum_{x=1}^L S_x^z$ to the Hamiltonian (\ref{Hop}).
We have computed the magnetization curve numerically for $L \le 32$
(as one example, the magnetization curve for $J=J'$ is shown in the inset
of Fig.\ \ref{mpdFig}) and then extrapolated it to the thermodynamic limit.
The final result is shown in Fig.\ \ref{mpdFig}. First, we find an
$\langle M \rangle = 0$ plateau which is equivalent to the spin
gap (accordingly, its boundary curve in Fig.\ \ref{mpdFig} is equivalent to
Fig.\ \ref{gapAF}).
In addition, we find a clear plateau at
$\langle M \rangle = 1/2$ (in a normalization where the fully polarized
state has $\langle M \rangle = 1$). Such a plateau is expected
on the basis of the limit of decoupled quadrumers (\ref{Hquadru}) --
see \cite{mRev} and references therein.
Similar behavior has been observed for the related model of ref.\
\cite{RIS} where an interesting sequence of magnetization plateaux
was found \cite{KOK}. One remarkable feature of the present model is
that there is evidence for a transition in the $\langle M \rangle = 1/2$
plateau state at $J \approx 0.65 J'$. 

\begin{thebibliography}{99}
\bibitem{footnTrSe} Besides the excitation considered
              in the present section, there is another low-lying excitation
              which becomes the lowest excited state for $J_2/2 < J' < J_2$
              and $J_1 = 0$ (see Table \ref{TabQspec}). Perturbative
              expansions for both excitations were performed numerically in 
              \cite{TrSe}.
\bibitem{footn2} We were unable to reproduce the best fit of \cite{GVAHW}:
              $\delta = 0.2$, $J \approx 126$meV. Indeed, it was pointed out
              in \cite{GVAHW} that the cluster-operator theory on which this
              fit was based is not accurate in the region of interest.
\end{thebibliography}
\end{document}